\documentclass[acmtog,nonacm,screen]{acmart}
\acmSubmissionID{1133}

\usepackage{booktabs} 
\usepackage{hyperref}
\usepackage[capitalize]{cleveref}

\usepackage[ruled]{algorithm2e} 
\usepackage{bm}
\usepackage{mathtools}
\usepackage{graphicx}
\usepackage{booktabs}
\usepackage{multirow}
\usepackage{tabularx}
\usepackage{enumitem}
\usepackage{tabularx}
\usepackage{xcolor}
\usepackage[most]{tcolorbox}

\SetAlFnt{\small}
\SetAlCapFnt{\small}
\SetAlCapNameFnt{\small}
\SetAlCapHSkip{0pt}

\acmJournal{TOG}

\begin{document}

\title{VoxScene: Anchor-Conditioned Voxel Diffusion for Indoor Scene Arrangement}

\author{Haotian Mao}
\orcid{0009-0005-1294-4337}
\email{maohaotian@sjtu.edu.cn}
\author{Yuhan Huang}
\email{huang_yuhan@sjtu.edu.cn}

\author{Jiatao Lin}
\email{1169652042@sjtu.edu.cn}
\author{Yang Zhao}
\orcid{0009-0009-9841-5813}
\email{runder1103@sjtu.edu.cn}

\author{Hui Wang}
\affiliation{%
 \institution{Shanghai Jiao Tong University}
 \city{Shanghai}
 \country{China}}
\email{wanghehv@sjtu.edu.cn}
\orcid{0000-0002-4554-0719}

\author{Yiheng Zhang}
\affiliation{%
 \institution{Hong Kong University of Science and Technology}
 \city{Hong Kong}
 \country{China}}
\email{e1349382@u.nus.edu}
\orcid{0009-0000-2889-4470}

\author{Yuwang Wang}
\email{wang-yuwang@tsinghua.edu.cn}
\affiliation{%
 \institution{Tsinghua University}
 \city{Beijing}
 \country{China}}
\orcid{0000-0002-6880-959X}

\author{Chenliang Zhou}
\email{chenliang.zhou@qq.com}
\affiliation{%
 \institution{University of Cambridge}
 \country{United Kingdom}}

\author{Yan Zhang}
\affiliation{%
 \institution{Shanghai Jiao Tong University}
 \city{Shanghai}
 \country{China}}
\email{zhangyan20086930@gmail.com}

\author{Fangcheng Zhong}
\affiliation{%
 \institution{Peking University}
 \country{China}}
\email{zfc@pku.edu.cn}

\author{Xubo Yang}
\affiliation{%
 \institution{Shanghai Jiao Tong University}
 \country{China}}
\email{yangxubo@sjtu.edu.cn}

\begin{abstract}
We present VoxScene, a novel anchor-conditioned voxel diffusion framework tailored for 3D scene synthesis. Current data-driven layout generation techniques typically rely on bounding proxies or implicit representations, which overlook volumetric structures. This geometric blindness inevitably leads to severe physical collisions and structural entanglement, particularly in densely populated environments. To overcome these limitations, we shift the paradigm to an explicit, object-centric voxel representation. Our pipeline sequentially synthesizes discrete volumetric occupancies conditioned on prior anchors and local context. By exploiting the mutually exclusive nature of discrete voxels, our approach eliminates spatial ambiguities and guarantees collision-free arrangements, even in highly complex environments. Furthermore, the synthesized high-fidelity voxel grids serve as discriminative geometric queries for downstream asset retrieval. Extensive experiments demonstrate the universality of our method, achieving state-of-the-art physical plausibility and unlocking shape diversity compared to existing layout planners.
\end{abstract}

%
%


%
%


\begin{teaserfigure}
    \centering
    \includegraphics[width=\textwidth]{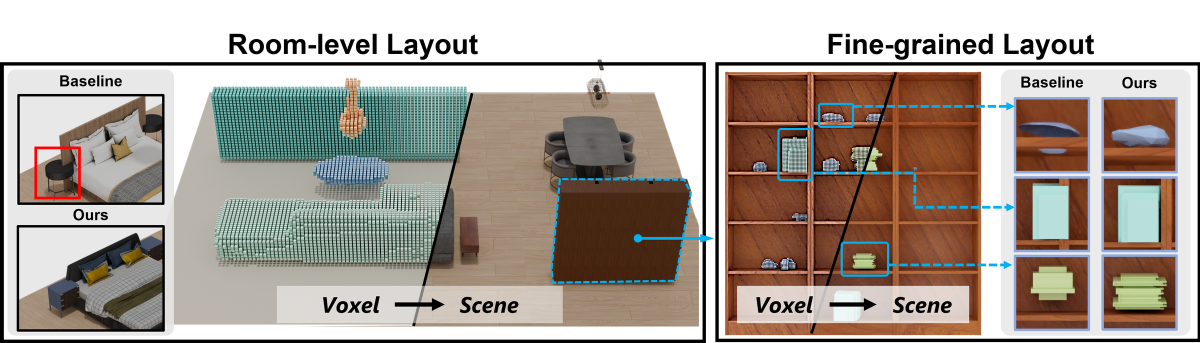}
    \caption{\textbf{VoxScene} introduces a voxel-based layout representation. By leveraging the spatial exclusivity of discrete occupancies, our method resolves object intersections (shown in the \textcolor{red}{red} box). Furthermore, the explicit 3D structure of voxels excels in handling fine-grained layouts within complex geometric contexts (shown in the \textcolor{blue}{blue} boxes). }
    \label{fig:teaser} 
\end{teaserfigure}

\maketitle

\section{Introduction}

3D indoor scene synthesis has drawn growing interest because of its extensive applications in virtual asset creation \cite{maleki2024procedural, wang2025large}, interior design \cite{liu2025computer}, and embodied AI \cite{duan2022survey, liu2025aligning}. By automating the placement and instantiation of furniture, these generative techniques significantly accelerate prototype development and reduce manual labor. With the development of deep neural networks, generative models, such as generative adversarial networks (GANs) \cite{nauata2020house, nauata2021house, tang2023graph}, autoregressive models \cite{vaswani2017attention,esser2021taming}, and diffusion models \cite{ho2020denoising,rombach2022high}, have significantly boosted layout diversity. More recently, large language models (LLMs) have enabled more flexible natural-language control for 3D generation \cite{feng2023layoutgpt,yangHolodeckLanguageGuided2024,ccelen2024design,sunLayoutVLMDifferentiableOptimization2025,yangSceneWeaverAllinOne3D2025,fuAnyHomeOpenvocabularyGeneration2025,yang2025OptiSceneLLMdrivenIndoor}, leveraging their multi-modal reasoning abilities to improve the generalization of existing frameworks.

Despite recent progress, synthesizing physically plausible dense scenes remains a formidable challenge. Existing paradigms share a fundamental flaw: they lack a direct, explicit perception of 3D physical boundaries. Without such volume awareness, they inevitably suffer from severe spatial mismatches, instance entanglement, and physical collisions. Initial work simplifies the layout problem by representing objects as bounding boxes \cite{paschalidouATISSAutoregressiveTransformers2021,linInstructSceneInstructiondriven3D2023}, which overlooks the fine-grained shape details. Current state-of-the-art techniques typically employ implicit formats, such as latent features \cite{tangDiffuSceneDenoisingDiffusion2024,yangPhyScenePhysicallyInteractable2024,liu2023openshape} and signed distance fields (SDFs) \cite{wei2025Planner3DLLMenhancedGraph,bokhovkinSceneFactorFactoredLatent,meng2025LT3SDLatentTrees}, to model intricate geometries. However, these methods lack explicit geometric constraints, limiting their ability to prevent inter-object penetration. Although LLM-based systems leverage open-vocabulary capabilities and multi-modal visual feedback to alleviate placement errors \cite{feng2023layoutgpt,sunLayoutVLMDifferentiableOptimization2025}, they remain bottlenecked by an absence of native 3D spatial understanding. This deficiency is critically magnified when handling complex spatial structural relations and densely populated arrangements such as nested and stacked configurations.

To overcome these limitations, we propose VoxScene (Shown in \cref{fig:teaser}), a novel voxel diffusion framework that shifts the scene modeling paradigm from ambiguous abstractions to an explicit, discrete volumetric representation. We formulate our task as an anchor-conditioned synthesis problem, positioning VoxScene as a highly robust generative backend that refines readily available spatial priors from upstream methods into detailed and physically plausible 3D arrangements. By operating directly on discrete voxel grids, our method naturally enforces hard occupancy constraints, substantially reducing inter-object penetrations and improving structural validity. To achieve high-fidelity volumetric structures while maintaining computational tractability during training, we design an object-centric diffusion strategy, which sequentially instantiates objects conditioned on their localized spatial context. This design effectively preserves instance-level boundaries and fine geometric details. Moreover, the explicit formulation allows our framework to better respect complex structural constraints, enabling dense object arrangements. Extensive experiments on both room-level and shelf-level synthesis demonstrate that VoxScene outperforms baselines in quantitative metrics. Conditioned on various upstream anchors from different sources, such as LLMs or other data-driven predictions, our method consistently enhances the physical plausibility (red box in \cref{fig:teaser}). In particular, existing methods often break down under complex structural constraints, while VoxScene remains stable and produces plausible arrangements (blue box in \cref{fig:teaser}). In addition, the explicit volumetric representation also promotes richer shape diversity, leading to greater geometric and structural variation in synthesized scenes. 

Our main contributions can be summarized as follows:
\begin{itemize}
\item We introduce a novel explicit volumetric representation paradigm for 3D scene layout synthesis. Its exclusive voxel occupancy design enforces non-overlapping spatial allocation at the layout level.
\item We design an object-centric voxel diffusion framework conditioned on prior anchors. By sequentially instantiating objects within their localized spatial context, our framework rigorously preserves instance-level boundaries and fine-grained volumetric details.
\item We conduct extensive experiments to validate the effectiveness and flexibility of our approach in both room-level tasks and delicate object arrangements. VoxScene can serve as a plug-and-play generative backend compatible with diverse upstream anchor generators, achieving superior physical plausibility and shape diversity.
\end{itemize}

\section{Related Work}

\subsection{Indoor Scene Synthesis}

Generating plausible indoor layouts has been widely explored in computer graphics. Traditional procedural approaches \cite{raistrickInfinigenIndoorsPhotorealistic, deitkeProcTHORLargeScaleEmbodied,yuMakeItHome} synthesize large-scale scenes using hand-crafted rules or heuristics, but require substantial manual design. Recent LLM-based approaches \cite{zhou2024SceneXProceduralControllable, chenSceneFoundryGeneratingInteractive2026, sun20253d, hu2024scenecraft} enable natural-language interaction and open-vocabulary control, yet their coverage is still limited by underlying heuristic rules. With the availability of massive 3D datasets, such as 3D-FRONT \cite{fu20213d}, Matterport3D\cite{chang2017matterport3d}, Structured3D \cite{zheng2020structured3d} and M3DLayout \cite{zhang2025m3dlayout}, data-driven generative models have made substantial progress. Existing methods employ diverse architectures, such as generative adversarial networks (GANs), variational auto-encoders (VAEs), autoregressive models\cite{paschalidouATISSAutoregressiveTransformers2021,wang2021SceneFormerIndoorScene, fengCASAGPTCuboidArrangement2025, bucher2025ReSpaceTextDriven3D}, masked non-autoregressive methods \cite{choi2026SceneNATMaskedGenerative} and diffusion models \cite{tangDiffuSceneDenoisingDiffusion2024, hu2024MixedDiffusion3D, linInstructSceneInstructiondriven3D2023, zhaiEchoSceneIndoorScene2025, zhai2023CommonScenesGeneratingCommonsense, yangFlowSceneStyleconsistentIndoor2026, yangPhyScenePhysicallyInteractable2024}. Another line of work lifts 2D priors to 3D scenes via distillation \cite{zhouGALA3DTextto3DComplex2024, gaoGraphDreamerCompositional3D2024}, multi-view generation \cite{nguyenHouseCrafterLiftingFloorplans2024, li2025WorldGrowGeneratingInfinite, fang2025SPATIALGENLayoutguided3D}, and projection-based techniques \cite{fuAnyHomeOpenvocabularyGeneration2025}. More recently, multi-modal LLM agentic systems have emerged for scene arrangement \cite{yangSceneWeaverAllinOne3D2025,pfaff2026SceneSmithAgenticGeneration, xiaSAGEScalableAgentic2026}, which utilize the planning capabilities of LLMs to iteratively invoke various tools to synthesize comprehensive layouts. Despite impressive results, their long tool invocation chains and reliance on physics engines make them time-consuming and difficult to scale for batch generation.

\begin{figure*}[th] 
    \centering
    \includegraphics[width=\linewidth]{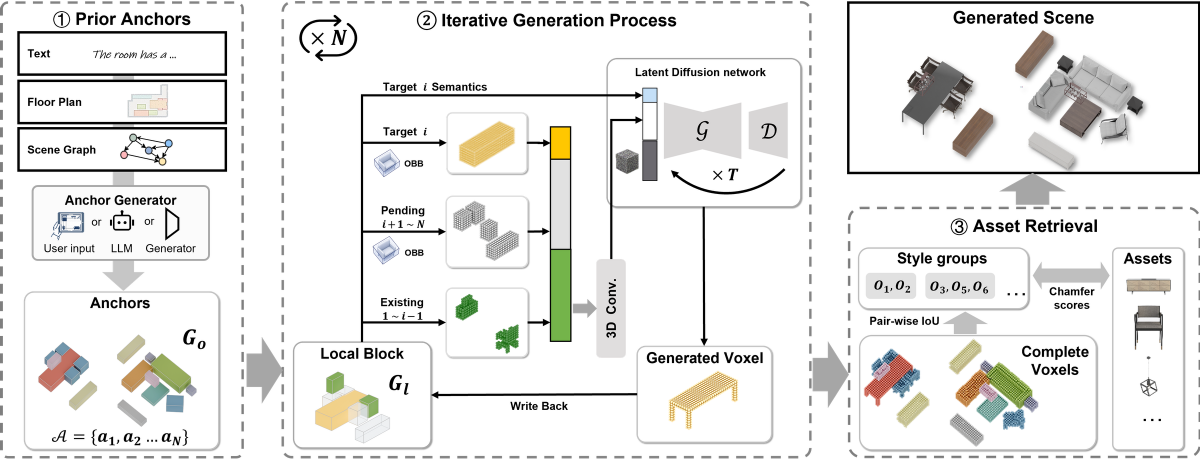}
    \caption{\textbf{Method overview}. We formulate scene synthesis as an anchor-conditioned generative process. Guided by prior anchors from various upstream sources (Part 1), our object-centric framework sequentially generates each target object, completing the whole scene represented in voxels through N iterations (Part 2). These generated voxels then serve as explicit geometric proxies for downstream asset retrieval, ultimately yielding a high-fidelity scene (Part 3).}
    \label{fig:pipeline} 
\end{figure*}

\subsection{Object Representation in Scene Synthesis}
Object representation fundamentally dictates the editability and physical plausibility of synthesized scenes, particularly serving as a crucial basis for downstream object retrieval. Early methods, such as ATISS \cite{paschalidouATISSAutoregressiveTransformers2021}, parameterize objects with bounding box attributes. While effective for coarse spatial planning, this formulation overlooks fine-grained geometry, significantly restricting the diversity of generated scenes. Subsequent works encode 3D objects into implicit shape codes \cite{tangDiffuSceneDenoisingDiffusion2024} using pre-trained models \cite{liu2023openshape,yang2018foldingnet}, facilitating their integration into transformer-based networks. Although this latent representation improves the network's generalization, optimizing layouts within the latent space inherently discards explicit boundaries. Consequently, decoding these representations often leads to severe mesh interpenetration. While some physics-guided frameworks like PhyScene \cite{yangPhyScenePhysicallyInteractable2024} incorporate physical constraints to mitigate this issue, they still struggle with out-of-bound errors and collisions, primarily due to the geometric mismatch between proxy representations and actual meshes. Recently, CASAGPT \cite{fengCASAGPTCuboidArrangement2025} decomposes 3D meshes into explicit cuboid primitives for autoregressive modeling to avoid object intersections. However, this abstraction sacrifices fine-grained geometric details such as curved surfaces or irregular shapes. In contrast, our method uses voxel grids to represent the precise spatial occupancy, which drives the subsequent retrieval process. By capitalizing on the discrete, non-overlapping nature of voxels, our pipeline guarantees more reasonable arrangements and scene synthesis.

\subsection{Voxel Diffusion for Scene Synthesis}
Voxel-based diffusion models have demonstrated remarkable success in 3D generation. While pioneering works like XCube \cite{ren2024XCubeLargeScale3D, xiang2025structured,sella2023voxe,ding2025FullPartGeneratingEach} utilize voxels for high-fidelity, object-level synthesis, scaling these discrete representations to complex scene-level generation remains challenging. Current scene-level diffusion frameworks primarily focus on capturing the holistic room layout. Works like BlockFusion \cite{wu2024BlockFusionExpandable3D}, SceneFactor \cite{bokhovkinSceneFactorFactoredLatent}, and Lt3sd \cite{xiang2025structured} model the entire scene as a monolithic global Signed Distance Field (SDF). While these global representations effectively model the overall spatial envelope, they fuse all furniture and architectural elements into a single continuous structure. Such techniques overlook instance-level isolation, significantly complicating downstream applications that require structural independence, such as object retrieval and interactive scene editing. Moreover, generating the entire scene in a single shot often sacrifices local details, degrading object-level geometry. Accordingly, our work introduces a context-aware object-centric pipeline, which explicitly maintains instance boundaries while maintaining fine-grained local geometries. 

\section{Method}

We introduce VoxScene, a novel generative paradigm that reformulates 3D scene layout synthesis through explicit voxel representations. As depicted in \cref{fig:pipeline}, our pipeline is driven by an anchor-conditioned, object-centric voxel diffusion model. Here, an \textbf{anchor} serves as a spatial condition to indicate the initial placement of an object. To preserve high-quality geometric details, our model instantiates objects sequentially. By conditioning the denoising network on previously synthesized objects and pending surrounding anchors as local spatial context, our approach ensures geometric consistency and collision avoidance. Ultimately, the generated voxel grids serve as robust geometric queries for retrieving high-quality 3D assets, completing the scene assembly.

\begin{figure*}[th] 
    \centering
    \includegraphics[width=\linewidth]{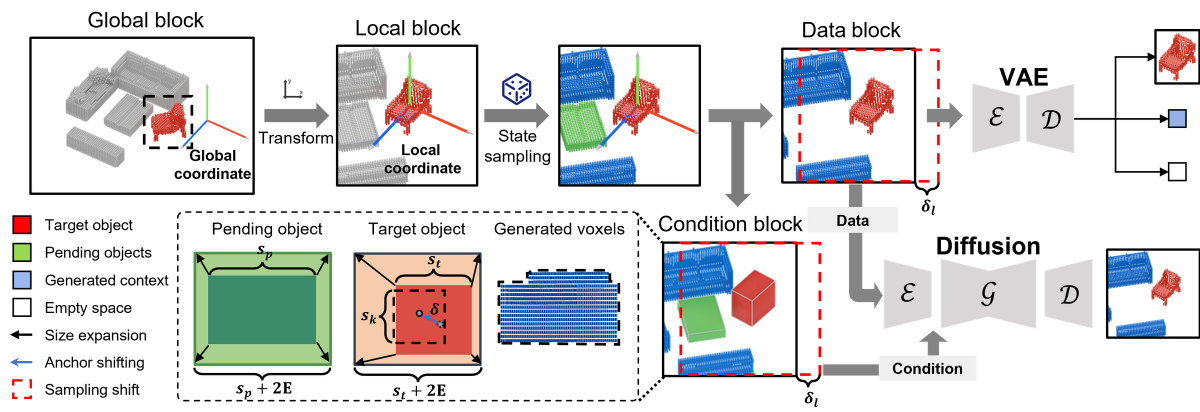}
    \caption{\textbf{Training policy}. We employ a stochastic masking policy to enable our model to generate in an arbitrary sequence. The anchor shift policy further improves the robustness and effectiveness.}
    \label{fig:training_policy} 
\end{figure*}

\subsection{Anchor-conditioned Denoising Process}

\textbf{Prior Anchor.} Formally, let $\mathcal{A} = \{\mathbf{a}_i\}_{i=1}^N$ denote the set of spatial anchors, where $N$ is the number of target objects. Each anchor is parameterized as $\mathbf{a}_i = \{\mathbf{p}_i, \mathbf{r}_i, \mathbf{s}_i, \mathbf{y}_i\}$, encapsulating the spatial position $\mathbf{p}_i \in \mathbb{R}^3$, the continuous 2D heading orientation $\mathbf{r}_i \in \mathbb{R}^2$, the geometric size $\mathbf{s}_i \in \mathbb{R}^3$, and a semantic category vector $\mathbf{y}_i \in \{0, 1\}^{N_C}$, where $N_C$ denotes the total number of object classes. Conditioned on $\mathcal{A}$ and a structural constraint $S$, our core objective is to synthesize a physically plausible and geometrically accurate layout, explicitly represented as a voxel grid where each occupied voxel is assigned a distinct instance identifier $i \in \{1,\dots, N\}$.

\textbf{Object-centric Generation.} However, a monolithic voxel representation inherently lacks the structural isolation required to distinguish individual instances. Furthermore, this global approach severely compromises fine geometric details while imposing a prohibitive memory overhead. To overcome these limitations, we formulate the global scene synthesis as a sequential, target-centric generative process. For each anchor $\mathbf{a}_i$, we perform synthesis within a local block $G_l$ of fixed resolution $K^3$, which is centered at $\mathbf{p}_i$ and aligned with the heading orientation $\mathbf{r}_i$. Since global IDs are arbitrarily assigned and completely lack cross-scene consistency, directly feeding them into the neural network would severely disrupt the learning process. To address this, we remap these absolute identifiers into a relative state $\tau \in \{0, 1, 2\}$, corresponding to free space ($\tau=0$), contextual objects ($\tau=1$), and the target object ($\tau=2$), respectively. Once generated, the voxels classified as $\tau=2$ are inversely transformed and written back into the global grid $G_o$. This operation is exclusively restricted to the remaining free space in $G_o$, preventing any modification to pre-existing non-empty voxels, which ensures that previously established geometries are rigorously preserved throughout the iterative process.

\textbf{Latent Space Compression.} Although our object-centric formulation reduces the spatial scope, modeling the generative process directly on raw voxel grids remains computationally prohibitive. Therefore, we leverage a 3D fully convolutional Vector-Quantized Variational Autoencoder (VQ-VAE) to compress the local voxel grids by a factor of 4. Unlike standard continuous VAEs, which often blur sharp geometric boundaries due to their continuous Gaussian priors, the discrete codebook of VQ-VAE perfectly aligns with the categorical nature of our ternary states. Specifically, the complete local input volume $\mathbf{x}$, including the discrete ternary states $x_{State}$ and the continuous distance fields $x_{SDF}$, is quantized into a compressed discrete latent $\mathbf{z}_q$ through an encoder $E$. We employ a shared decoder backbone $D$ equipped with a primary state head $H_{\text{state}}$ and an auxiliary geometry head $H_{\text{SDF}}$. These components are jointly optimized using the following objective:
\begin{equation}
\begin{split}
    \mathcal{L} &= \mathcal{L}_{CE}(\mathbf{x}, H_{state}(D(\mathbf{z}_q))) + \\ & \quad \mathcal{L}_{quant}(E(\mathbf{x}), \mathbf{z}_q) + \mathcal{L}_1(\mathbf{x}, H_{SDF}(D(\mathbf{z}_q))), 
\end{split}
\end{equation}
where $\mathcal{L}_{\text{CE}}$ represents the cross-entropy loss for the state channels $x_{state}$, and $\mathcal{L}_{\text{quant}}$ denotes the standard codebook quantization and commitment loss to align the continuous encoder outputs $E(x)$ with the discrete latent codes $\mathbf{z}_q$. Notably, we independently compute the ground-truth SDF for the isolated target object as supervision for $x_{SDF}$, effectively guiding the network to better capture the geometric structures.

\textbf{Anchor Conditions.} During the sequential generative process $1 \dots N$, the conditioning signal at step $i$ is defined by three disjoint subsets: the already generated and structural voxels $\mathcal{V}_{\text{gen}} = \{\mathbf{v}_1, \dots, \mathbf{v}_{i-1},\mathbf{v}_{S}\}$, the current target anchor $\mathbf{a}_i$, and the pending anchors $\mathcal{A}_{\text{p}} = \{\mathbf{a}_{i+1}, \dots, \mathbf{a}_N\}$. As illustrated in Part 2 of \cref{fig:pipeline}, to ensure stable and precise spatial control, these geometric conditions are rasterized into dense voxel-wise condition tensors, denoted as $\mathbf{c}_{\text{gen}}$, $\mathbf{c}_{\text{t}}$, and $\mathbf{c}_{\text{p}}$, respectively.

The generated context $\mathcal{V}_{gen}$ is rasterized into an $N_c$-channel tensor using its voxel occupancies and semantic labels. Since geometries of pending objects are unavailable, their anchors are converted to coarse Oriented Bounding Boxes (OBBs) and voxelized into another $N_c$-channel tensor. Overlapping OBBs are merged by a logical OR operation, producing a multi-hot semantic encoding.

For the target object condition $c_t$, we explicitly decouple its semantic and spatial signals. The semantic cue is given by the target category vector $\mathbf{y}_i$, which specifies the object class. The spatial condition is a 2-channel tensor: One channel encodes the target OBB to restrict the generative region, while the other encodes an anisotropic volumetric kernel of size $\mathbf{s}_k \in \mathbb{Z}^3$ to indicate the synthesis origin. Compared with sparse point signals, this volumetric kernel is more robust to down-sampling and supports the anchor shifting strategy in \cref{sec:training}. 

All spatial condition tensors ($\mathbf{c}_{\text{gen}}$, $\mathbf{c}_{\text{p}}$, and the spatial part of $\mathbf{c}_{\text{t}}$) are processed through a 3D convolutional down-sampling network to the VQ-VAE latent resolution and concatenated with the noisy latent features. The semantic vector $\mathbf{y}_i$ is broadcast across the latent space and concatenated as global category conditioning.

\begin{figure}[th] 
    \centering
    \includegraphics[width=\linewidth]{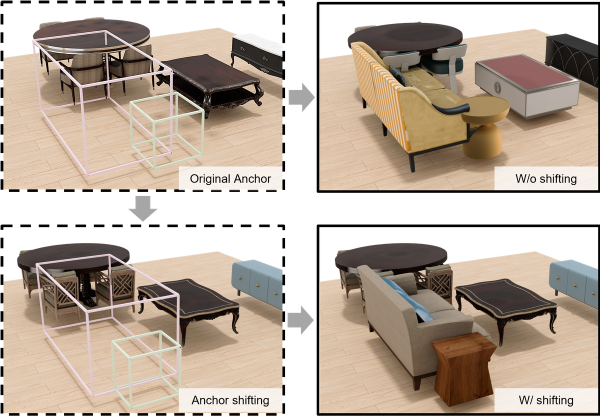}
    \caption{\textbf{Effect of anchor shifting}.}
    \label{fig:shifting} 
\end{figure}

\textbf{Denoising process.} Operating within the compact latent space of the VQ-VAE, we formulate the generative process as a Latent Diffusion Model (LDM). Let $\mathbf{z}_0$ denote the clean latent feature extracted from the VQ-VAE encoder. In the forward pass, $\mathbf{z}_0$ is progressively corrupted by injecting Gaussian noise $\mathbf{\epsilon} \sim \mathcal{N}(\mathbf{0}, \mathbf{I})$ over $T$ discrete timesteps until it reaches a nearly pure isotropic Gaussian distribution:
\begin{equation}
q(\mathbf{z}_t | \mathbf{z}_{t-1}) = \mathcal{N}(\mathbf{z}_t; \sqrt{1-\beta_t}\mathbf{z}_{t-1}, \beta_t \mathbf{I}),
\end{equation}
where $\beta_t$ is the variance schedule. Applying the reparameterization trick, the closed-form distribution of the noisy latent at any arbitrary timestep $t$ can be directly sampled without iterative computation:
\begin{equation}
q(\mathbf{z}_t | \mathbf{z}_0) = \mathcal{N}(\mathbf{z}_t; \sqrt{\bar{\alpha}_t}\mathbf{z}_0, (1-\bar{\alpha}_t)\mathbf{I}),
\end{equation}
where $\alpha_t \coloneqq 1 - \beta_t$ and $\bar{\alpha}_t \coloneqq \prod_{s=1}^t \alpha_s$ represents the cumulative signal retention.

To achieve superior geometric fidelity and faster convergence, our 3D U-Net $\mathbf{u}_\theta$ predicts the velocity target $\mathbf{u}_t$. Mathematically, the target velocity $\mathbf{u}_t$ is defined as the linear combination of the underlying noise and the clean latent data:
\begin{equation}
\mathbf{u}_t \coloneqq \sqrt{\bar{\alpha}_t} \bm{\epsilon} - \sqrt{1-\bar{\alpha}_t} \mathbf{z}_0.
\end{equation}

During the reverse denoising process, the network $\mathbf{v}_\theta$ takes the noisy latent $\mathbf{z}_t$, the timestep $t$, and the overall spatial condition $\mathbf{c}=\{c_{gen}, c_p,c_t\}$ as inputs. The overall latent diffusion objective is thus optimized via the mean squared error against the ground-truth velocity:
\begin{equation}
\mathcal{L}_{\text{LDM}} = \mathbb{E}_{\mathbf{z}_0, \bm{\epsilon} \sim \mathcal{N}(\mathbf{0}, \mathbf{I}), t} \left[ \left\| \bm{u}_\theta(\mathbf{z}_t, t, \mathbf{c}) - \mathbf{u}_t \right\|_2^2 \right].
\end{equation}

\subsection{Training Policy} \label{sec:training}
The joint training policy for our generative pipeline, encompassing stochastic context masking and anchor shifting, is illustrated in \cref{fig:training_policy}.

\textbf{Stochastic Context Masking.} To empower the model to synthesize scenes in an arbitrary order, we introduce a stochastic context masking strategy during the training phase. For each surrounding object within the local data block, we independently apply a dropout probability (empirically set to $p=0.5$) to determine its generation state. Based on this probability, a context object is randomly partitioned into either the generated set $\mathcal{V}_{\text{gen}}$ or the pending set $\mathcal{A}_{\text{p}}$.  This randomized masking encourages the network to adapt to diverse context densities and arbitrary generation sequences encountered during inference.

\textbf{Anchor shifting.} In practical inference scenarios, upstream anchors may suffer from positional noise or geometric overlap. We perturb the target anchor kernel during training by applying a random translation $\bm{\delta}$, sampled uniformly from $[-\mathbf{E}, \mathbf{E}] \cap \mathbb{Z}^3$. Accordingly, we set the anchor kernel size to $\mathbf{s}_k = 2\mathbf{E}+1$, ensuring that the original center remains covered under any perturbation. The extents of the target and pending anchors, originally $\mathbf{s}_t$ and $\mathbf{s}_p$, are expanded to $\mathbf{s}_t + 2\mathbf{E}$ and $\mathbf{s}_p + 2\mathbf{E}$, respectively. This expansion guarantees that the newly formed OBB fully encloses the original object geometry and provides sufficient space for voxel instantiation. Furthermore, to avoid a trivial center-bias where the network assumes the target is always aligned with the center of $G_l$, we independently shift the local grid origin by $\bm{\delta}_l$, sampled uniformly from $[-\mathbf{E}, \mathbf{E}]$. This strategy allows our model to fix the errors in prior anchors (shown in \cref{fig:shifting}).

\begin{figure}[!htpb] 
    \centering
    \includegraphics[width=\linewidth]{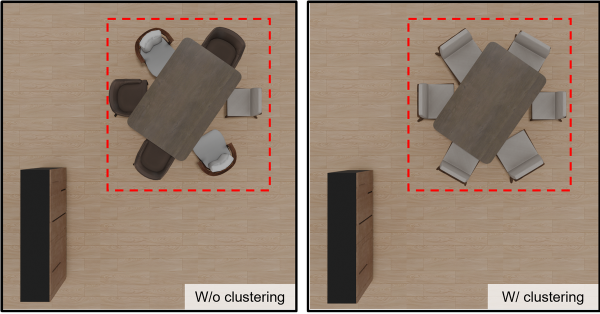}
    \caption{\textbf{Effect of style clustering}.}
    \label{fig:style_clustering} 
\end{figure}
\subsection{Anisotropy Asset Retrieval.}
\begin{figure*}[th] 
    \centering
    \includegraphics[width=\linewidth]{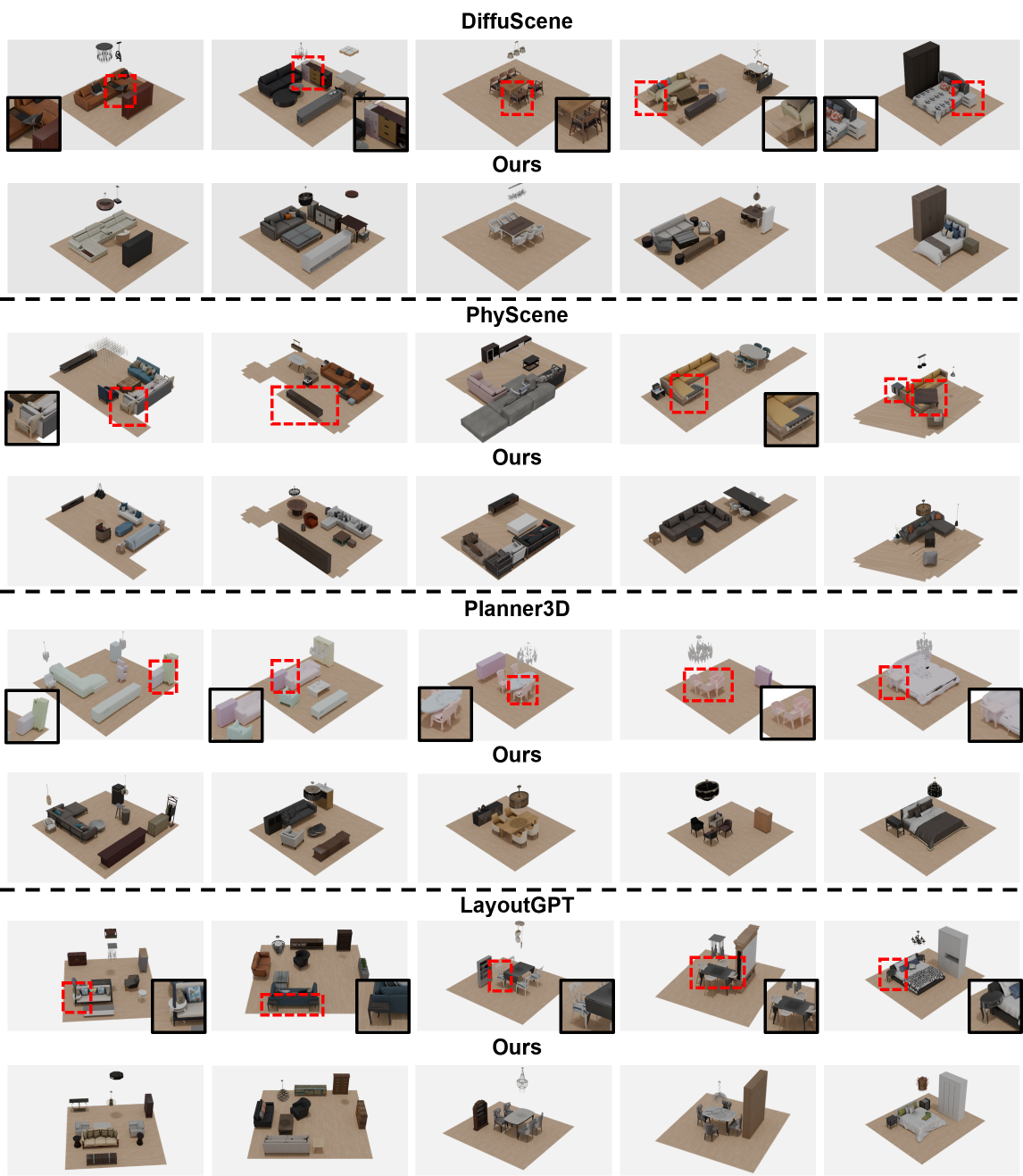}
    \caption{\textbf{Qualitative comparisons in 3D-FRONT}.}
    \Description{Qualitative comparisons in 3D-FRONT.}
    \label{fig:results1} 
\end{figure*}

\begin{figure*}[th] 
    \centering
    \includegraphics[width=0.95\linewidth]{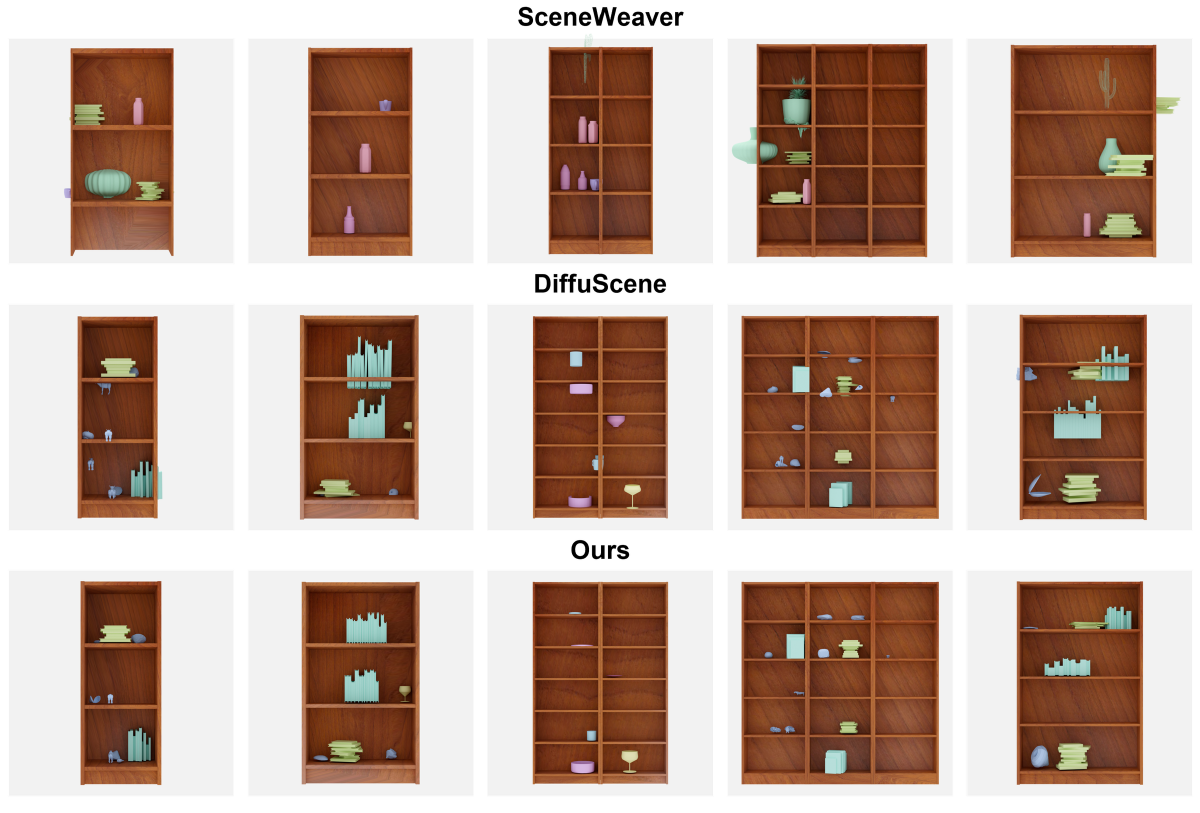}
    \caption{\textbf{Qualitative comparisons in M3D-Shelf.} Our model is generated conditioned on the anchors generated by DiffuScene. We chose the most similar scene from SceneWevaer as a comparison.}
    \label{fig:results2} 
\end{figure*}

\begin{figure*}[th] 
    \centering
    \includegraphics[width=\linewidth]{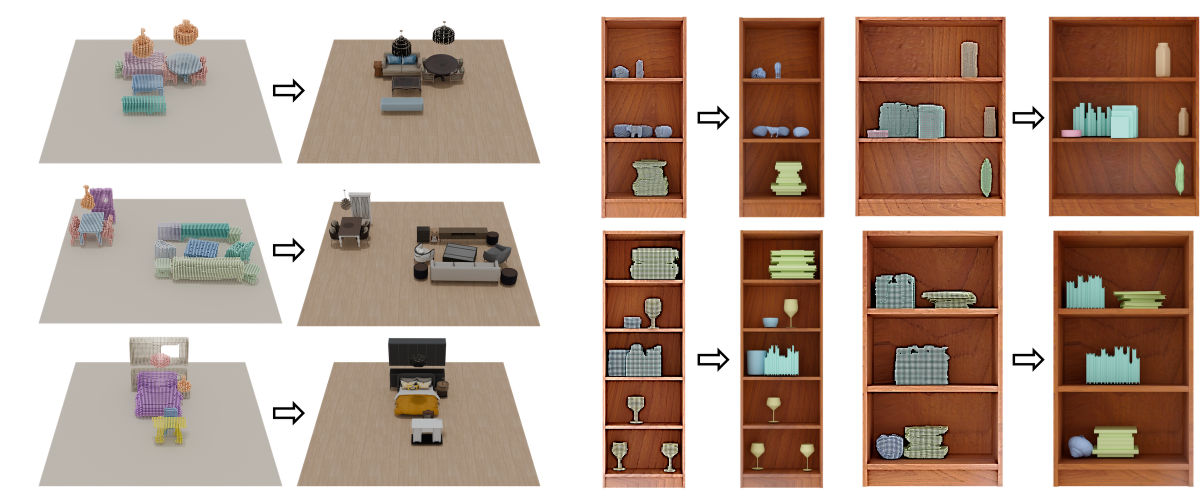}
    \caption{\textbf{More Results of our voxel generation and retrieved assets.}}
    \label{fig:vox_to_scene} 
\end{figure*}

Once the iterative diffusion process synthesizes the entire scene layout, we instantiate each generated voxel object with a realistic model. All models are pre-normalized into a canonical unit coordinate system and voxelized at a fixed resolution of $K_A^3$. During inference, the local block is sampled exactly at the anchor's center without the spatial perturbation. 

For each generated object, we extract its raw voxel occupancy and apply an inverse transformation matrix $R_v$ to map it into the canonical unit space.  To ensure geometric plausibility and prevent excessive structural distortion, we implement an anisotropic scaling threshold $t_a$. Specifically, while we allow independent scaling along the three local axes to fit the generated bounding box, any candidate asset that requires a scaling deformation exceeding $t_a$ is discarded from the pool. 

For the remaining valid candidates with the identical semantic label, we identify the optimal geometric match by computing the Soft Chamfer Score. Let $Q=\{q_i\}_{i=1}^{|Q|}$ and $C=\{c_j\}_{j=1}^{|C|}$ denote the voxel sets of the generated query and the candidate shapes, respectively. The unidirectional point-to-set distances are defined as:
\begin{equation}
  d(q_i, C)=\min_{c_j\in C}\|q_i-c_j\|_2,
  \quad
  d(c_j, Q)=\min_{q_i\in Q}\|c_j-q_i\|_2.
\end{equation}

To robustly evaluate the structural similarity against voxel discretization noise, the Soft Chamfer Score is formulated as a symmetric Gaussian-weighted average:

\begin{equation}
  S(Q,C) = \frac{1}{2} \left( \frac{1}{|Q|} \sum_{q_i\in Q} e^{-\frac{d(q_i,C)^2}{2\sigma^2}} + \frac{1}{|C|} \sum_{c_j\in C} e^{-\frac{d(c_j,Q)^2}{2\sigma^2}} \right),
\end{equation}
where $\sigma$ is the Gaussian tolerance in voxel units, we select the candidate asset $M_{1}$ that yields the maximum score $S(Q, C)$ and rigidly transform it back into the global scene coordinate to accomplish the final high-fidelity layout instantiation.

\begin{table*}[t]
    \centering
    \caption{Quantitative results on 3D-FRONT. The best results are marked in \textbf{bold}. The symbol $\downarrow$ indicates that lower values are better.}
    
    \small
    
    \renewcommand{\arraystretch}{1.2}
    
    \newcolumntype{Y}{>{\centering\arraybackslash}X}
    
    \begin{tabularx}{\textwidth}{@{} l *{12}{Y} @{}}
        \toprule
        \multirow{2}{*}{Methods} & \multicolumn{4}{c}{Bedroom} & \multicolumn{4}{c}{Livingroom} & \multicolumn{4}{c}{Diningroom} \\
        \cmidrule(lr){2-5} \cmidrule(lr){6-9} \cmidrule(lr){10-13}
        & $FID\downarrow$ & $I_p \downarrow$ & $OR\downarrow$ & $CD$
        & $FID\downarrow$ & $I_p \downarrow$ & $OR\downarrow$ & $CD$
        & $FID\downarrow$ & $I_p \downarrow$ & $OR\downarrow$ & $CD$ \\
        \midrule
        
        DiffuScene 
        & \textbf{25.37} & 8.53 & 20.45 & 0.522
        & 41.77 & 2.34 & 8.35 & 0.545
        & 38.23 & 2.51 & 19.16 & 0.555 \\
        DiffuScene + Ours
        & 27.79 & \textbf{3.89} & \textbf{7.08} & \textbf{0.715}
        & \textbf{38.83} & \textbf{1.89} & \textbf{7.18} & \textbf{0.911}
        & \textbf{36.88} & \textbf{1.65} & \textbf{11.92} & \textbf{0.842} \\
        \midrule 
        
        Planner3D 
        & 111.67 & 18.98 & 12.35 & -
        & 113.67 & 7.42 & 22.09 & -
        & 116.27 & 8.20 & 19.10 & - \\
        Planner3D + Ours
        & \textbf{74.25} & \textbf{5.67} & \textbf{7.79} & 0.276
        & \textbf{77.39} & \textbf{0.70} & \textbf{0.81} & 0.428
        & \textbf{80.21} & \textbf{1.18} & \textbf{4.33} & 0.382 \\
        \midrule 
        
        LayoutGPT
        & 73.66 & 4.00 & 1.26 & 0.210
        & \textbf{81.31} & 3.98 & 6.79 & 0.337
        & \textbf{86.37} & 0.82 & 1.79 & \textbf{0.318} \\
        LayoutGPT + Ours
        & \textbf{72.69} & \textbf{2.93} & \textbf{0.08} & \textbf{0.264}
        & 85.00 & \textbf{0.98} & \textbf{0.36} & \textbf{0.339}
        & 87.44 & \textbf{0.42} & \textbf{0.40} & 0.296 \\
        \bottomrule 
    \end{tabularx}
    \label{tab:result_3dfront}
\end{table*}

\begin{table}[tb]
    \centering
    \caption{Quantitative results on the \textit{Livingroom} compared with PhyScene. The best results are marked in \textbf{bold}. The symbol $\downarrow$ indicates that lower values are better.}
    \label{tab:physcene_standalone}
    
    \small 
    \renewcommand{\arraystretch}{1.1}
    
    \begin{tabular*}{\linewidth}{@{\extracolsep{\fill}} l cccccc @{}}
        \toprule
        Methods & $FID\downarrow$ & $I_p\downarrow$ & $OR\downarrow$ & $CD$ & $R_{o}\downarrow$ & $R_{vo}\downarrow$ \\
        \midrule
        
        PhyScene
        & 45.25 & 3.30 & 13.36 & 0.292 & 8.05 & 19.82 \\
        PhyScene + Ours 
        & \textbf{38.00} & \textbf{0.69} & \textbf{4.93}  & \textbf{0.990} & \textbf{4.01} & \textbf{13.81} \\
        \bottomrule
    \end{tabular*}
\end{table}

\textbf{Style Clustering.} To improve stylistic consistency, we propose a style clustering policy, unifying generated instances with the same semantic label. Specifically, two instances are connected if their aligned voxel occupancy IoU exceeds $t_{\text{IoU}}$ and their axis-wise size ratio lies within $t_s$, and each connected component defines a style cluster. During the retrieval stage, we select the representative instance with the highest Soft Chamfer Score and assign its top-1 retrieved CAD model to all instances in the cluster. This strategy effectively eliminates stylistic variations and enhances the overall visual coherence of the final instantiated scene (shown in \cref{fig:style_clustering}).

\section{Experiment}

\subsection{Experimental Setup}
\textbf{Datasets.} We evaluate the superiority of VoxScene on two distinct datasets spanning different spatial scales: 3D-FRONT \cite{fu20213d} for room-level layout generation and M3D-Shelf \cite{zhang2025m3dlayout} for fine-grained object arrangement. 3D-FRONT is a large-scale benchmark dataset for indoor scene synthesis. While M3D-Shelf is a customized subset extracted from M3DLayout, concentrating on delicate object arrangement in the shelf region.

\textbf{Baseline and Upstream Integration.} To demonstrate the universality and superiority of VoxScene, we evaluate our pipeline in conjunction with four layout generation frameworks: DiffuScene \cite{tangDiffuSceneDenoisingDiffusion2024}, PhyScene \cite{yangPhyScenePhysicallyInteractable2024}, Planner3D \cite{wei2025Planner3DLLMenhancedGraph}, and Layoutgpt \cite{feng2023layoutgpt} in 3D-FRONT evaluation. Particularly, these selected methods cover different types of input, including text prompts, floor plans, and scene graphs. In our experimental setup, these frameworks fulfill a dual role: First, their native layout predictions act as our primary competitive \textbf{baselines}. In the meanwhile, they function as upstream \textbf{anchor generators} for our method: we extract their predicted coarse spatial anchors and inject them as conditioning signals into our object-centric voxel diffusion model.

In M3D-Shelf, we choose DiffuScene and SceneWeaver \cite{yangSceneWeaverAllinOne3D2025} as our baselines, with DiffuScene also serving as our anchor generator. To accomplish this fine-grained object arrangement, we adapt it into a completion task where the shelf serves as the known condition for DiffuScene. For SceneWeaver, we execute its complete room-level generation pipeline and extract the shelf region for comparison.

For asset retrieval, we utilize the 3D-FUTURE dataset as our primary CAD database. Since Planner3D natively synthesizes geometries via SDFs, we preserve its original mesh extraction process. This respects the native geometric shapes, ensuring an optimal performance in physical plausibility.

\subsection{Evaluation on 3D-FRONT}

\textbf{Evaluation Metrics}
To comprehensively assess the generated scene layouts, we employ a suite of metrics focusing on statistical realism, physical plausibility, and shape diversity. Fréchet Inception Distance (FID) \cite{heusel2017gans} assesses overall distribution quality. To quantify the collision avoidance of our method, we introduce the average pairwise intersection $I_p$ and  $OR$.  And $OR$ evaluates the total percentage of intersecting voxels across the entire scene space. Out-of-bound object ($R_o$) and voxel ($R_{vo}$) ratios are measured with 1-voxel tolerance in comparison with PhyScene. Regarding diversity, category-weighted diversity $CD$ calculates the utilization rate of unique CAD assets retrieved from the dataset. Note that $I_p$ and $OR$ values are scaled by a factor of $10^3$.

\textbf{Result.} As demonstrated in \cref{tab:result_3dfront} and \cref{tab:physcene_standalone}, our explicit volumetric representation excels in physical plausibility while maintaining a competitive visual quality. Previous generative methods rely on object latent representations (DiffuScene and PhyScene) or implicit SDF representation (Planner3D), which inevitably introduce unexpected collisions. Moreover, Planner3D reconstructs independent SDFs without surrounding context, leading to more severe spatial intersections. And Layoutgpt heavily prioritize 2D visual layout patterns but fundamentally lack 3D spatial understanding, leaving significant room for geometric refinement. In contrast, VoxScene effectively resolves these spatial ambiguities. The non-overlapping nature of discrete voxel occupancies guarantees a more plausible arrangement. Regardless of the quality of the initial anchors, our method can greatly improve the collision situation while maintaining a competitive realism, even when conditioned on severely colliding spatial priors from Planner3D. The visualization comparisons are shown in \cref{fig:results1}.

Furthermore, VoxScene significantly enhances generation \textbf{diversity}. Existing methods frequently suffer from mode collapse, repeatedly retrieving only a narrow subset of the available CAD models. Instead, our pipeline successfully explores a much broader range. 

\begin{table}[b]
    \centering
    \caption{Quantitative results of different methods. Note that $R_F$ and OR are scaled by $10^3$. The best results are marked in \textbf{bold}.}
    \label{tab:m3dlayout}
    
    \small 
    \setlength{\tabcolsep}{3pt} 
    
    \begin{tabular}{l ccccccc}
        \toprule
        \multirow{2}{*}{Methods} & \multicolumn{7}{c}{Evaluation Metrics} \\
        \cmidrule(lr){2-8}
        & $R_o\downarrow$  & $R_f\downarrow$ & $R_s\downarrow$ & OR$\downarrow$ & $S_r$ & $S_l$ & $S_c$ \\
        \midrule
        DiffuScene       & 0.408 & 0.830 & 0.580 & 30.60  & 4.715 & 3.565 & \textbf{5.965} \\
        SceneWeaver      & 0.398 & 0.214 & 0.415 & 8.60  & 5.854 & 5.000 & 5.177 \\
        DiffuScene+Ours  & \textbf{0.080} & \textbf{0.184} & \textbf{0.034} & \textbf{3.37}  & \textbf{6.275} & \textbf{5.950} & 5.900 \\
        \bottomrule
    \end{tabular}
\end{table}

\subsection{Evaluation on M3D-Shelf}
\textbf{Evaluation Metrics.} 
For physical evaluation, we assess the ratio of out-of-bounds objects $R_o$, floating objects $R_f$, and shelf-colliding objects $R_s$. We also retain the global overlap ratio $OR$ in the previous experiment. Besides, we also employ GPT-5.4 as an automated evaluator to score the quality of realism ($S_r$), layout ($S_l$), and completion ($S_c$).

\textbf{Results.} As demonstrated in \cref{tab:m3dlayout}, VoxScene significantly outperforms the baseline methods across multiple dimensions. Illustrated in \cref{fig:results2}, although DiffuScene performs reasonably on room-level layouts, it fails to model fine-grained structural constraints, leading to collapsed arrangements with severe geometric conflicts. Similarly, while SceneWeaver employs iterative rule-based optimization to refine arrangements and somewhat mitigates these errors, it still leaves a substantial number of unresolved collisions. In contrast, VoxScene robustly respects the structural constraints and produces plausible arrangements even when conditioned on noisy anchors generated by DiffuScene. This correctness translates into superior perceptual quality, proving our method's exceptional capability in generating high-quality arrangements within complicated structures, specifically tailored for delicate and dense object configurations.

Beyond physical plausibility, our method demonstrates a decisive advantage in computational efficiency. Sceneweaver requires an average time of \textbf{612.14} seconds to synthesize a shelf containing merely \textbf{5.49} objects. This immense latency stems from its reliance on multiple external tool invocations, wherein its rule-based procedural solver alone consumes \textbf{289.13} seconds. Conversely, our method completes the generation in only \textbf{43.40} seconds, despite handling a denser average objects of \textbf{8.09}. This highlights the potential of our generative method to function as a collision handler, bypassing the computational overhead of traditional iterative solvers.

\subsection{Ablation Study}

We conduct extensive ablation studies on the \textit{livingroom} split of the 3D-FRONT dataset, utilizing prior anchors generated by DiffuScene. Quantitative results are summarized in \cref{tab:ablation_study}.

Although the auxiliary SDF head is discarded during the final asset retrieval phase, its inclusion during training serves as a crucial geometric regularizer. Removing it degrades all metrics, indicating that implicit distance supervision is essential for learning well-formed geometric boundaries. Furthermore, without the anchor shifting strategy, the network fails to autonomously rectify the spatial deviations inherent in upstream priors, leading to a noticeable drop in overall layout plausibility. Removing the pending anchor conditions $\mathbf{c}_p$ deprives the global context of uninstantiated objects. This absence of spatial foresight impairs the golistic coordination of the scene layout. Particularly, our style clustering policy presents an intentional trade-off between visual quality and physical constraints. While it achieves aesthetic harmony, assigning a shared representative CAD model introduces slight geometric deviations compared to the optimally matched raw voxels.

\begin{table}[tb]
    \centering
    \caption{\textbf{Ablation study on different settings}. The best results are marked in \textbf{bold}. The symbol $\downarrow$ indicates that lower values are better.}
    \label{tab:ablation_study}
    
    \small 
    \setlength{\tabcolsep}{8pt} 
    
    \begin{tabular}{l ccc}
        \toprule
        {Settings} & $FID\downarrow$ & $I_p\downarrow$ & $OR\downarrow$ \\
        
        \midrule
        DiffuScene            & 41.77 & 2.34 & 8.35 \\
        w/o SDF head            & 41.78 & 2.94 & 10.44 \\
        w/o anchor shifting            & 41.20 & 2.52 & 8.66 \\
        w/o pending anchor conditions & 39.20 & 1.92 & 7.22 \\
        w/o style cluster            & 38.91 & \textbf{1.86} & \textbf{7.07} \\
        Full       & \textbf{38.83} & 1.89& 7.18 \\
        \bottomrule
    \end{tabular}
\end{table}

\section{Conclusions}

In this paper, we have introduced VoxScene, a novel explicit volumetric representation paradigm for 3D scene layout synthesis. Conditioned on coarse spatial anchors, our object-centric diffusion framework synthesizes detailed, physically plausible volumetric occupancies, demonstrating superior spatial reasoning in complex structures. Furthermore, VoxScene functions as a highly versatile generative backend that can be seamlessly integrated with diverse upstream semantic planners—ranging from data-driven models and LLM-based agents to direct user inputs—to significantly elevate their geometric fidelity.

\textbf{Limitations and Future Work.} Although our object-centric diffusion strategy robustly rectifies minor collisions, it is primarily designed for local geometric refinement rather than large-scale global repositioning. It may not correct the serious problem in prior anchors, such as objects that overlap completely. This can be improved through a rejection policy for abnormal voxels, such as low chamfer scores or obvious absence. Future work could explore joint optimization strategies that propagate spatial collision gradients from our retrieval algorithm back to the upstream planners, thereby enabling a fully end-to-end, geometry-aware scene generation pipeline.

%
%
%
%

\bibliographystyle{ACM-Reference-Format}
\bibliography{sample-bibliography}

\clearpage
\appendix

\section{Data Processing}

\subsection{Voxelization}
Given a 3D scene $R = \{M_1, M_2, ..., M_N, S\}$, composed of $N$ object meshes and a fixed architectural structure $S$, we first discretize the continuous scene into the global instance grid $G_o \in \mathbb{N}^{D \times W \times H}$ with voxel size $s_v$. To prevent slender geometries from degrading or fracturing during this discrete sampling, we determine voxel occupancies using a robust surface-band threshold $b$. Subsequently, a 3D morphological hole-filling algorithm is applied to solidify the internal cavities of the objects. Through this process, each object $M_i$ is converted to a corresponding dense volumetric occupancy grid $O_i$, which explicitly encodes its corresponding semantic category $\mathbf{y}_i$. Crucially, raw 3D assets often contain inherent interpenetrations, and the voxelization process itself can introduce quantization errors, occasionally causing a single voxel to be claimed by multiple instances. To resolve these spatial conflicts, we introduce a size-based priority heuristic: if a voxel's occupancy is contested, it is exclusively assigned to the object with the larger bounding volume. This rigorously enforces a mutually exclusive assignment constraint $O_i \cap O_j = \emptyset$, ensuring that any single voxel belongs to at most one specific instance, and is explicitly assigned a label $(\mathbf{y}_i, i)$ to denote its semantic category and instance identifier.

For each object $O_i$, we establish a local voxel grid $G_l^i$ with a fixed resolution of $K^3$ centered at its origin, maintaining the identical physical voxel size $s_v$ of the global space $G_o$. Considering that forcing the network to model arbitrary orientations imposes an immense learning burden and severely compromises fine details, we explicitly canonicalize each local volume by applying the inverse of the object's spatial rotation. To guarantee global contextual consistency, we populate the local grid $G_l^i$ by directly sampling the semantic and instance tuples from the global grid $G_o$ via nearest-neighbor interpolation. Consequently, each canonicalized local block encapsulates both the target object and its surrounding context information.

\section{Experiment Details} 

\subsection{Dataset}

\textbf{3D-FRONT.} To ensure a fair and rigorous comparison, we strictly adhere to the data filtering protocols established by DiffuScene \cite{tangDiffuSceneDenoisingDiffusion2024} in the 3D-FRONT experiment. The curated dataset comprises 4,041 bedrooms, 900 dining rooms, and 813 living rooms for training and testing. 

\textbf{M3D-Shelf.} M3DLayout is a high-quality indoor scene layout dataset that notably features the arrangement of delicate objects. To evaluate our model's capability in fine-grained spatial reasoning, we curate a customized subset denoted as M3D-Shelf. To build this subset, we exclusively extract the cabinet/shelf structures alongside their supported items. Before voxelization, we convert every shelf and contained object Blender/Z-up frame into our project Y-up coordinate system. For each asset, we fit the mesh to the local axis-aligned bounding box provided by M3DLayout, apply the layout \textit{matrix\_world}, and then center the resulting shelf scene. This curated subset comprises 490 shelf environments, with each containing 3 to 20 objects. We partition the data into training and testing splits using a standard 80\%/20\% ratio.

In M3D-Shelf, the shelf body is treated as a structural context rather
than as a generated furniture target. Concretely, the shelf body is assigned the
semantic name \textit{largeshelf}, voxelized into the structure occupancy, and
provided to both object-centric conditioning and downstream evaluation. The
contained objects are the generation targets. Corresponding semantic categories and rendering colors are listed in \cref{tab:supp_m3d_semantics_colors}.

\subsection{Implementation Details}
In our experiment, we configure different voxel sizes for each dataset: $0.0375m$ for 3D-FRONT and a finer $0.01m$ for M3D-Shelf, considering the varying spatial scales of our target domains. Across both experiments, the local bounding grid $G_l$ maintains a fixed spatial resolution of $64^3$ voxels. The style clustering strategy is exclusively applied to the 3D-FRONT dataset, employing a $t_{IoU}=0.3$  and $t_{s}=1.1$. Regarding the spatial perturbation strategy, the maximum translation ranges for the random shift range $\mathbf{E}$ are empirically set to $[4, 0, 4]$ and $[6, 6, 6]$ voxels along the three spatial axes for the two datasets, respectively.

We train individual object-centric diffusion models for each scene category. All experiments are conducted on two NVIDIA RTX 4090 GPUs. The VQ-VAE is trained for 300 epochs using the standard Adam optimizer \cite{kingma2014adam} with a constant learning rate of $2 \times 10^{-4}$. Subsequently, the diffusion models are optimized via AdamW \cite{loshchilov2017decoupled} with a learning rate of $1 \times 10^{-5}$ to ensure stable convergence and robust regularization. Specifically, to accommodate the distinct geometric complexities of the two tasks, the diffusion models undergo 450 training epochs for the 3D-FRONT dataset and 500 epochs for the fine-grained M3D-Shelf dataset.

\begin{table}[tb]
    \centering
    \caption{M3D-Shelf semantic categories and their corresponding evaluation colors. HEX values and visual swatches are provided for reference.}
    \label{tab:supp_m3d_semantics_colors}
    \small
    \renewcommand{\arraystretch}{1.1}
    
    \newcommand{\coloritem}[1]{\texttt{\##1} \quad \textcolor[HTML]{#1}{\rule{2.5em}{0.8em}}}

    \begin{tabular}{ll}
        \toprule
        Semantic category & HEX \& Visual Color \\
        \midrule
        \texttt{largeshelf} & \coloritem{7A8285} \\
        \texttt{bookcolumn} & \coloritem{FABD2E} \\
        \texttt{bookstack} & \coloritem{F29433} \\
        \texttt{bottle} & \coloritem{00A1B8} \\
        \texttt{bowl} & \coloritem{D17AC7} \\
        \texttt{can} & \coloritem{8C94A3} \\
        \texttt{cup} & \coloritem{3B7DF2} \\
        \texttt{foodbag} & \coloritem{DB6E57} \\
        \texttt{foodbox} & \coloritem{E6B24C} \\
        \texttt{hardware} & \coloritem{333842} \\
        \texttt{jar} & \coloritem{2EAD5C} \\
        \texttt{natureshelftrinkets} & \coloritem{B28040} \\
        \texttt{pan} & \coloritem{61616B} \\
        \texttt{plate} & \coloritem{F2EBC7} \\
        \texttt{pot} & \coloritem{A87052} \\
        \texttt{wineglass} & \coloritem{6BCCEB} \\
        \bottomrule
    \end{tabular}
\end{table}

\subsection{Baseline}

In the 3D-FRONT experiment, all comparative evaluations are conducted using the official open-source repositories of the respective baselines. For DiffuScene and PhyScene, we directly deploy their publicly released pre-trained weights. Because pre-trained weights were unavailable for Planner3D, we trained the model from scratch for 180 epochs across all room categories, strictly adhering to the configurations specified in their official documentation. Notably, since the official PhyScene repository only offers a checkpoint for the living rooms without the training code, our comparative analysis against this specific baseline is confined to the living room domain.

In the M3D-Shelf experiment, we follow the official setting for DiffuScene, training the autoencoder for 1000 epochs and the diffusion model for 35000 epochs. For SceneWeaver, we used five room-conditioned prompts, each specifying a compact indoor scene with a single open-front shelf. The prompts share the same structural constraint: the shelf must remain visibly open and must not be converted into a closed cabinet with doors, drawers, or solid panels (\cref{tab:supp_open_shelf_prompts}). During testing, we sample prompts from this set and pass the selected text to SceneWeaver to meet the user demand.

\begin{table*}[t]
\caption{Prompts and invocation protocol for the SceneWeaver open-shelf test.}
\label{tab:supp_open_shelf_prompts}
\centering
\small

\definecolor{AcBlue}{RGB}{41, 96, 143}     
\definecolor{AcBg}{RGB}{248, 250, 252}     

\begin{tcolorbox}[
    enhanced,
    colback=AcBg,              
    colframe=AcBlue,           
    coltitle=white,            
    title=\textbf{SceneWeaver Open-Shelf Prompts},
    fonttitle=\bfseries,       
    boxrule=0.8pt,             
    arc=3pt,                   
    auto outer arc,
    left=10pt, right=10pt, top=6pt, bottom=8pt, 
    toptitle=4pt, bottomtitle=4pt
]

\newcommand{\cleanprompt}[2]{
    \vspace{0.6em}\noindent\textbf{#1} #2
}

\cleanprompt{Office:}{Design a compact office with one open-front shelf. The shelf must stay visibly open, with no doors, drawers, or closed cabinet panels. Make the shelf look naturally used by arranging appropriate small items inside its shelves.}

\cleanprompt{Kitchen:}{Design a compact kitchen with one open-front shelf. The shelf must stay visibly open, with no doors, drawers, or closed cabinet panels. Make the shelf look naturally used by arranging appropriate small items inside its shelves.}

\cleanprompt{Living room:}{Design a compact living room with one open-front shelf. The shelf must stay visibly open, with no doors, drawers, or closed cabinet panels. Make the shelf look naturally used by arranging appropriate small items inside its shelves.}

\cleanprompt{Bedroom:}{Design a compact bedroom with one open-front shelf. The shelf must stay visibly open, with no doors, drawers, or closed cabinet panels. Make the shelf look naturally used by arranging appropriate small items inside its shelves.}

\cleanprompt{Bathroom:}{Design a compact bathroom with one open-front shelf. The shelf must stay visibly open, with no doors, drawers, or closed cabinet panels. Make the shelf look naturally used by arranging appropriate small items inside its shelves.}

\vspace{1em}
\hrule height 0.5pt \vspace{0.8em} 

\noindent\textbf{Invocation:} For each trial, SceneWeaver first initializes a compact room using the sampled user prompt and room type, then reloads the scene and adds only small objects inside the existing shelf, using the shelf as the parent relation target.

\end{tcolorbox}
\end{table*}

Unless otherwise specified, repeated SceneWeaver trials are produced by sampling
from these five prompts with replacement under the benchmark seed. No prompt is
manually adjusted for a particular generated scene. 

\subsection{Evaluation Metric Calculation Details}

All geometric metrics are computed after the retrieval stage using the placed CAD meshes. For each evaluated object mesh, we first build a surface voxel grid with pitch $s_e$, then apply 3D hole filling to obtain a solid occupied voxel set $V_i$. We choose $0.02m$ and $0.012m$ in two experiments, respectively.

\textbf{Fréchet Inception Distance (FID).} Following previous works, we render top-down orthographic images at a resolution of $256 \times 256$, coloring instantiated objects according to their semantic category. Let
$(\mu_r,\Sigma_r)$ and $(\mu_g,\Sigma_g)$ be the Gaussian statistics of
Inception features from real and generated renderings. We compute
\begin{equation}
\mathrm{FID} =
\|\mu_r-\mu_g\|_2^2 +
\mathrm{Tr}\left(\Sigma_r+\Sigma_g-2(\Sigma_r\Sigma_g)^{1/2}\right).
\end{equation}

\textbf{Average Pairwise Intersection ($I_p$).} For a scene with $N$ objects,
let $\mathcal{P}=\{(i,j)\mid 1\leq i<j\leq N\}$ be all object pairs, and let
$V_i$ be the solid occupied voxel set of object $i$. The pairwise intersection
severity is
\begin{equation}
s_{ij} =
\max\left(
\frac{|V_i\cap V_j|}{|V_i|},
\frac{|V_i\cap V_j|}{|V_j|}
\right).
\end{equation}
Then $I_p$ is calculated as:
\begin{equation}
I_p=\frac{1}{|\mathcal{P}|}\sum_{(i,j)\in\mathcal{P}}s_{ij}.
\end{equation}

\textbf{Global Overlap Ratio (OR).} OR measures the total intersected object volume
relative to non-overlapped occupied volume:
\begin{equation}
\mathrm{OR} =
\frac{\sum_{(i,j)\in\mathcal{P}} |V_i\cap V_j|}
{\sum_{i=1}^{N}|V_i|-\sum_{(i,j)\in\mathcal{P}} |V_i\cap V_j|+\epsilon}.
\end{equation}

\textbf{Floor-plan violation ($R_o$ and $R_{vo}$).} For 3D-FRONT-style room scenes, let
$F\subset\mathbb{R}^2$ be the valid floor-plan region on the horizontal $xz$
plane, and let $\pi_{xz}$ project a voxel center to this plane. To keep the
notation consistent with the intersection metrics above, we use the same solid
occupied voxel set $V_i$ for object $i$. With a boundary tolerance $\eta$, the
out-of-floor voxels of object $i$ are
\begin{equation}
V_i^{\mathrm{out}}=
\{\mathbf{x}\in V_i\mid \pi_{xz}(\mathbf{x})\notin F^{+\eta}\},
\end{equation}
where $F^{+\eta}$ denotes the floor-plan region expanded by $\eta$. We set
$\eta=0.02m$ in our 3D-Front experiment. The object-level violation rate is
\begin{equation}
R_o=\frac{1}{N}\sum_{i=1}^{N}
\mathbf{1}\left[|V_i^{\mathrm{out}}|>0\right],
\end{equation}
and the out-of-bound voxel ratio is
\begin{equation}
R_{vo}=
\frac{\sum_{i=1}^{N}|V_i^{\mathrm{out}}|}
{\sum_{i=1}^{N}|V_i|}.
\end{equation}
$R_o$ measures how many objects cross the floor-plan boundary, while $R_{vo}$
measures how much occupied volume lies outside the valid floor region.
\begin{table}[!htbp]
    \centering
    \caption{Training resource usage of the M3D-Shelf VQ-VAE and object-centric diffusion models (measured on two NVIDIA RTX 4090 GPUs).}
    \label{tab:supp_training_resource}
    \small
    \renewcommand{\arraystretch}{1.1} 
    
    \begin{tabular*}{\linewidth}{@{\extracolsep{\fill}} l c c r @{}}
        \toprule
        Model (Setting) & Batch & Peak Mem. & Time/Step \\
        \midrule
        
        VQ-VAE 
        & \multirow{2}{*}{4} & \multirow{2}{*}{8.1 GiB} & \multirow{2}{*}{0.12 s} \\
        \footnotesize{\textit{($64^3$ local blocks)}} & & & \\
        
        \addlinespace 
        
        Diffusion 
        & \multirow{2}{*}{4} & \multirow{2}{*}{12.8 GiB} & \multirow{2}{*}{0.45 s} \\
        \footnotesize{\textit{($16^3$ latents + online)}} & & & \\
        
        \bottomrule
    \end{tabular*}
\end{table}

\begin{table}[!htbp] 
    \centering
    \caption{Average runtime of our M3D-Shelf generation pipeline (evaluated on a single NVIDIA RTX 4090 GPU over 200 scenes containing 1,617 objects).}
    \label{tab:supp_runtime_summary}
    \small
    \renewcommand{\arraystretch}{1.2} 
    
    \begin{tabular*}{\linewidth}{@{\extracolsep{\fill}} ll r r @{}}
        \toprule
        Pipeline Stage & Execution Setting & Time / Scene & Time / Object \\
        \midrule
        Diffusion Generation & Single-card & 23.15 s & 2.86 s \\
        CAD Retrieval        & Single-thread & 20.25 s & 2.50 s \\
        \midrule
        \textbf{End-to-end}  & \textbf{Sequential} & \textbf{43.41 s} & \textbf{5.37 s} \\
        \bottomrule
    \end{tabular*}
    
    \vspace{0.4em}
    \begin{minipage}{\linewidth}
        \footnotesize \textit{Note:} Diffusion time is derived from a single-card sequential estimate. Retrieval time is extrapolated from a single-thread sample. Per-object averages are amortized over the 1,617 generated objects.
    \end{minipage}
\end{table}

\begin{table}[!htbp]
    \centering
    \caption{Average SceneWeaver runtime for shelf population only, amortized per inserted object.}
    \label{tab:supp_sceneweaver_runtime}
    \small
    \renewcommand{\arraystretch}{1.2} 
    
    \begin{tabular*}{\linewidth}{@{\extracolsep{\fill}} l r r @{}}
        \toprule
        Pipeline Stage & Time / Scene & Time / Object \\
        \midrule
        GPT object planning         & 60.39 s & 10.61 s \\
        Load initialized scene      & 48.69 s & 8.56 s \\
        Graph action                & 10.48 s & 1.84 s \\
        Pre-solve asset population  & 123.66 s & 21.73 s \\
        Shelf-placement solver core & 63.07 s & 11.08 s \\
        Solve and cleanup           & 226.06 s & 39.73 s \\
        Record scene                & 253.99 s & 44.64 s \\
        \midrule
        \textbf{Shelf-population total} & \textbf{612.14 s} & \textbf{107.58 s} \\
        \bottomrule
    \end{tabular*}
\end{table}

\textbf{M3D-Shelf spatial validity.} Let $\Omega_S$ denote the aligned valid
shelf volume, with the front opening treated as the only permissive side. Using
the same solid occupied voxel set $V_i$ as above for object $i$, we define the
out-of-shelf voxels
\begin{equation}
V_i^{\mathrm{out}}=\{\mathbf{x}\in V_i \mid \mathbf{x}\notin \Omega_S^{+\eta}\},
\end{equation}
where $\Omega_S^{+\eta}$ is the shelf volume. An object is counted as out-of-bounds if $V_i^{\mathrm{out}}$ is nonempty,
except for a benign spill through the $z_{\max}$ opening whose projection remains
inside the opening aperture and does not violate the other faces. The
out-of-bounds rate is
\begin{equation}
R_o=\frac{1}{N}\sum_{i=1}^{N}\mathbf{1}\!\left[|V_i^{\mathrm{out}}|>0\right].
\end{equation}
This metric has the same object-level meaning as the floor-plan $R_o$, but the
calculation is different: the floor-plan version checks a 2D room boundary after
projecting voxels to the $xz$ plane, while the M3D-Shelf version checks a 3D
shelf volume with a special opening-face exemption.

The shelf-collision rate is
\begin{equation}
R_s=\frac{1}{N}\sum_{i=1}^{N}\mathbf{1}\!\left[\mathcal{T}_S\cap B_i^{-}\neq\emptyset\right],
\end{equation}
where $\mathcal{T}_S$ is the set of shelf mesh triangles and $B_i^{-}$ is the
object axis-aligned bounding box (AABB) shrunk by the intrusion margin. 

The floating rate is
\begin{equation}
R_f=\frac{1}{N}\sum_{i=1}^{N}\mathbf{1}\!\left[h_i^{\mathrm{sup}}\ \text{is undefined}\ \vee\ h_i^{\mathrm{bottom}}-h_i^{\mathrm{sup}}>\delta\right],
\end{equation}
where $h_i^{\mathrm{bottom}}$ is the object bottom height and $h_i^{\mathrm{sup}}$
is the support height under the object center. OR uses the
same object-object formulas above, computed only over contained objects.

We set the tolerance margin to ${0.012m}$ for the open face and ${0.036m}$ for the remaining shelf boundaries.

\textbf{LLM assessment.} We render one front-facing semantic-color
view for each M3D-Shelf scene (following \cref{tab:supp_m3d_semantics_colors}) and provide the image, semantic color mapping, and object metadata to the evaluator. The LLM assigns integer scores
$g_{n,k}\in\{0,\ldots,10\}$ for realism, layout, and completion. The evaluation prompt is shown in \cref{tab:supp_llm_eval_prompt}, for which we made minimum modification from SceneWeaver to adapt it to our shelf-level tasks.

\begin{table*}[t]
\caption{Prompt and scoring protocol for the M3D-Shelf LLM assessment.}
\label{tab:supp_llm_eval_prompt}
\centering
\small

\definecolor{AcBlue}{RGB}{41, 96, 143}
\definecolor{AcBg}{RGB}{248, 250, 252}

\begin{tcolorbox}[
    enhanced,
    colback=AcBg,
    colframe=AcBlue,
    coltitle=white,
    title=\textbf{M3D-Shelf LLM Evaluation Prompt},
    fonttitle=\bfseries,
    boxrule=0.8pt,
    arc=3pt,
    auto outer arc,
    left=10pt, right=10pt, top=6pt, bottom=8pt,
    toptitle=4pt, bottomtitle=4pt
]

\newcommand{\evalprompt}[2]{
    \vspace{0.55em}\noindent\textbf{#1} #2
}

\evalprompt{Task:}{You are given a single \texttt{\{llm\_view\}} render of an extracted open-front shelf/cabinet. Evaluate how well the objects are placed inside or on the shelf. This is a cabinet-level evaluation, not a whole-room evaluation; ignore the rest of the room.}

\evalprompt{Rendering:}{The image uses semantic colors rather than realistic textures. Ignore texture, lighting, and image style. Use geometry, object categories, positions, sizes, and visible placement quality.}

\evalprompt{Strict scoring:}{For each criterion, assign an integer score from 0 to 10 and provide a brief justification. Score 10 means the result fully meets the criterion, score 5 indicates noticeable flaws, and score 0 indicates that the aspect is absent, wrong, or contradicts the request. Critical issues such as floating objects, objects outside the shelf, hidden objects behind closed panels, severe collisions, implausible categories, or missing shelf contents should significantly lower the relevant score.}

\evalprompt{Realism:}{Assess whether the shelf/cabinet placement is physically plausible. Good results show a visibly open shelf/cabinet with plausible shelf objects, believable object sizes, and a naturally used arrangement. Bad results include closed or visually nonsensical shelves, implausible categories, severe scale errors, floating objects, or deeply embedded objects.}

\evalprompt{Layout:}{Assess whether objects are geometrically arranged correctly inside or on the shelf. Good results have objects supported by shelf surfaces, inside the shelf footprint, upright or reasonably oriented, not severely colliding, not outside the cabinet, and not blocking the front opening. Bad results contain floating objects, board penetration, heavy collisions, outside placement, buried objects, wrong scale/orientation, or objects invisible from the front.}

\evalprompt{Completion:}{Assess whether the result is sufficiently populated as an open-front shelf/cabinet. Good results contain an appropriate number and variety of small objects and look intentionally arranged. Bad results are empty, nearly empty, sparse, irrelevant, or do not meaningfully populate the shelf. Do not require whole-room furniture or room-level functionality.}

\evalprompt{Metadata:}{The prompt includes four JSON blocks: target shelf/cabinet metadata, contained/on-shelf object metadata, object semantic color mapping, and camera/render metadata. The metadata and image must be used together; objects should be counted carefully and checked for visibility, support, scale, placement, and severe collision.}

\evalprompt{Coordinate conventions:}{The metadata is in a right-handed local coordinate system with meter units. The $Y$ axis is up. The shelf front/opening in the rendered frame is \texttt{\{opening\_direction\}} (\texttt{\{opening\_face\_name\}}). Object local location is the center, size is given as \texttt{[x\_half, y\_half, z\_half]}, and AABBs are local-frame bounds for quick checking.}

\evalprompt{Return format:}{Return only a JSON object with three keys: \texttt{realism}, \texttt{layout}, and \texttt{completion}. Each key contains an integer \texttt{grade} and a short \texttt{comment}. If the image and metadata conflict, briefly mention the conflict in the relevant comment.}

\end{tcolorbox}
\end{table*}

\begin{figure*}[t] 
    \centering
    \includegraphics[width=\linewidth]{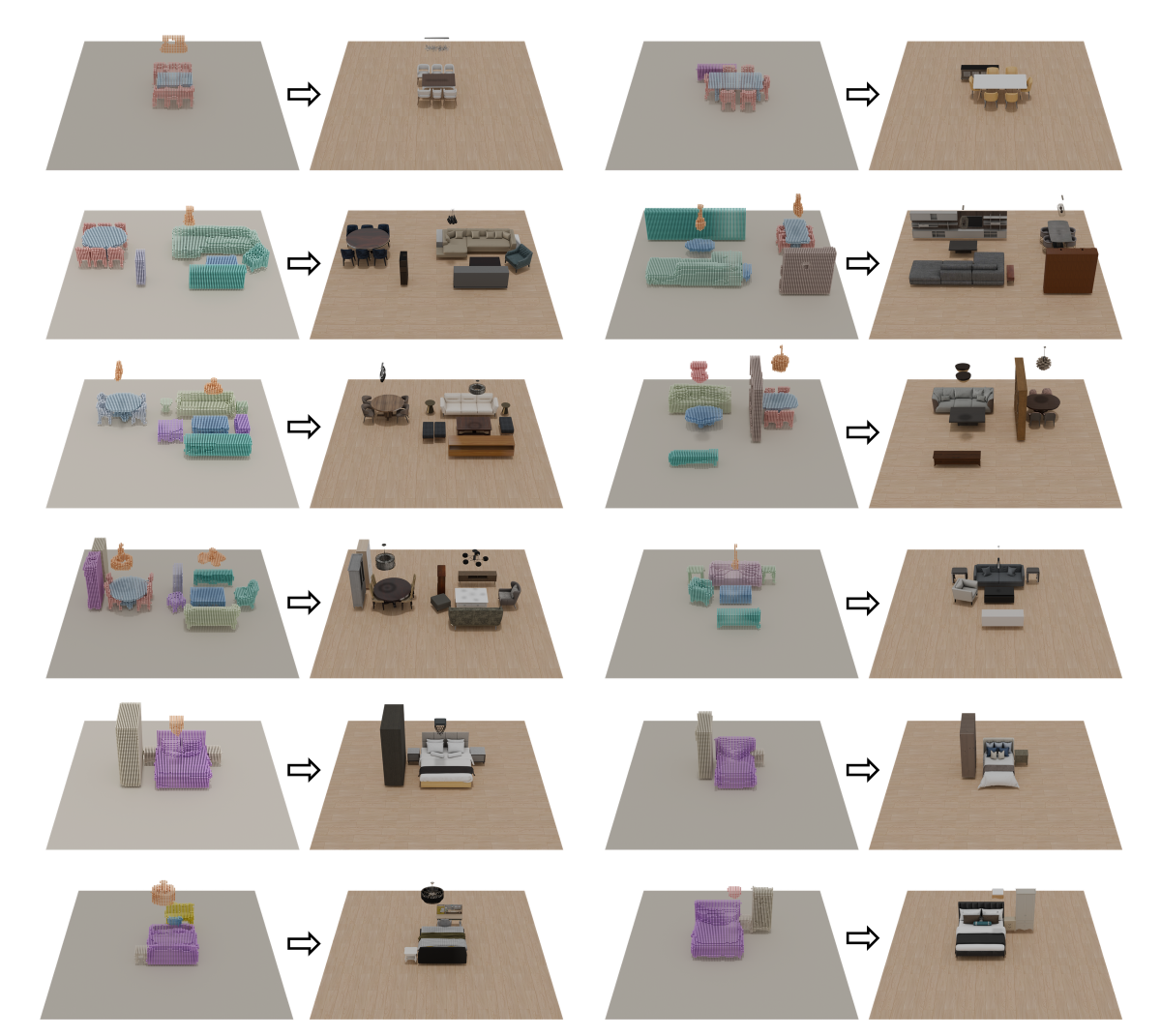}
    \caption{More results on 3D-FRONT.}
    \label{fig:mat_results1} 
\end{figure*}

\begin{figure*}[t] 
    \centering
    \includegraphics[width=\linewidth]{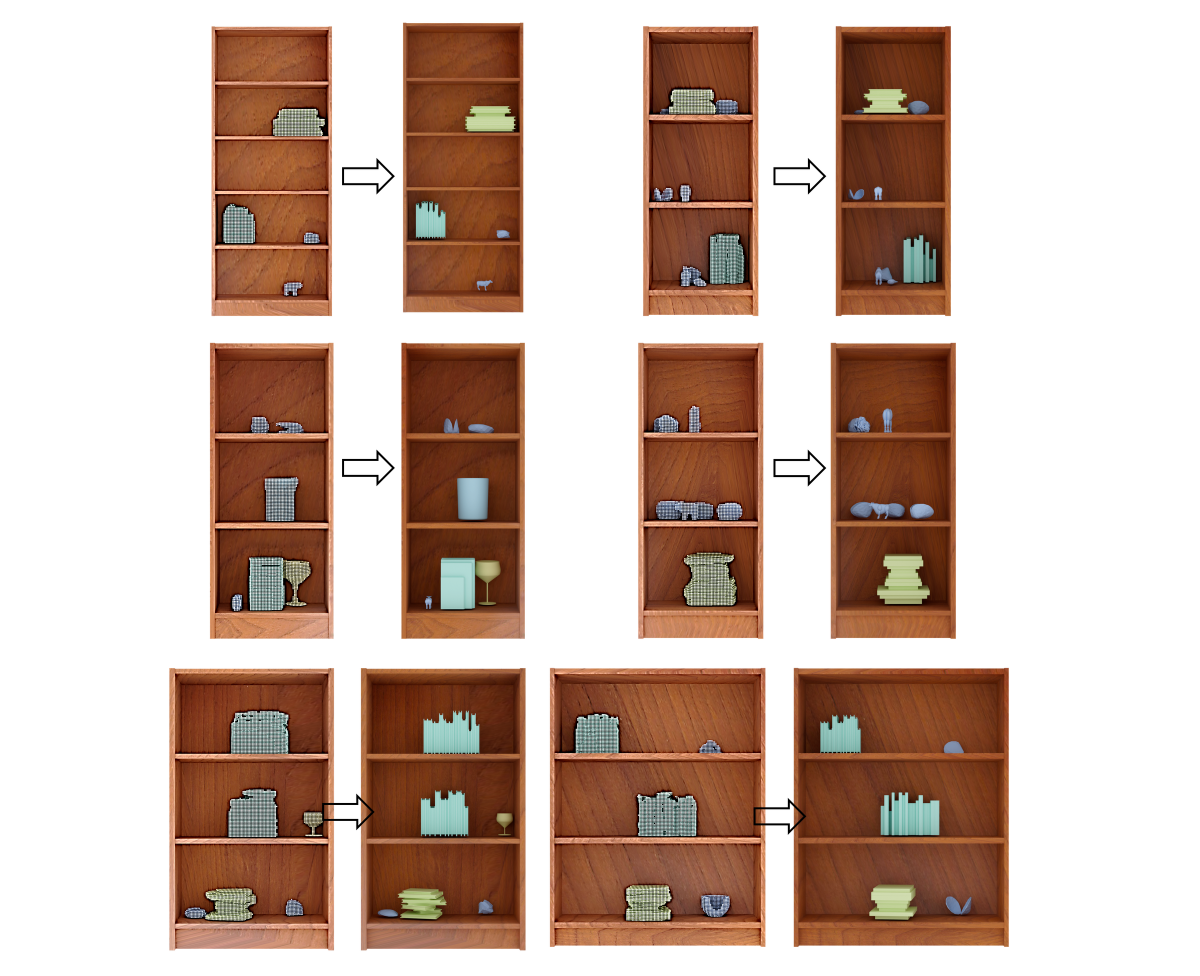}
    \caption{More results on M3D-Shelf.}
    \label{fig:material_results2} 
\end{figure*}

\section{Other Results}

\subsection{Memory Consumption}
We report our memory values in the training stage in \cref{tab:supp_training_resource}.

\subsection{Generation Efficiency}

The efficiency of our pipeline is reported in \cref{tab:supp_runtime_summary}. As a comparison, we report the average wall-clock time over the 200 successful SceneWeaver
open-shelf runs (\cref{tab:supp_sceneweaver_runtime}). Times are measured in seconds. The initialization stage creates
the compact room and open shelf, while the object-addition stage reloads the
scene and places small objects inside the shelf.

\end{document}